 \documentclass[preprint,superscriptaddress,showpacs,prl]{revtex4-1}
\usepackage[dvips]{graphicx}
\usepackage{epsfig}
\begin{document}

%\title[Impact of drift wave turbulence on the equilibrium distribution function]{Impact of drift wave turbulence on the equilibrium distribution function}
\title[A theory of non-local linear drift wave transport]{A theory of non-local linear drift wave transport}
\author{S. Moradi$^{1}$, J. Anderson$^{1}$ and B. Weyssow$^{2}${\it \small $^1$\\
Department of Applied Physics, Nuclear Engineering, Chalmers
  University of Technology and Euratom-VR Association, G\"oteborg,
  Sweden\\
$^{2}$EFDA-CSU, D-85748 Garching, M\"{u}nchen, Germany}}

%\author{B. Weyssow}
%\address{EFDA-CSU, D-85748 Garching, M\"{u}nchen, Germany}
%\ead{sara@nephy.chalmers.se}
\begin{abstract}
Transport events in turbulent tokamak plasmas often exhibit non-local or non-diffusive action at a distance features that so far have eluded a conclusive theoretical description. In this paper a theory of non-local transport is investigated through a Fokker-Planck equation with fractional velocity derivatives. A dispersion relation for density gradient driven linear drift modes is derived including the effects of the fractional velocity derivative in the Fokker-Planck equation. It is found that a small deviation (a few percent) from the Maxwellian distribution function alters the dispersion relation such that the growth rates are substantially increased and thereby may cause enhanced levels of transport.
\end{abstract}

\pacs{52.25.Dg,52.30.Gz,52.35.Kt}
\maketitle

\section{Introduction}
Understanding anomalous transport in magnetically confined plasmas is an outstanding issue in controlled fusion research. A satisfactorily understanding of the non-local features as well as the non-Gaussian probability distribution functions (PDFs) found in experimental measurements of particle and heat fluxes is still lacking. In particular, experimental observations of the edge turbulence in the fusion devices \cite{Zweben} show that in the Scrape of Layer (SOL) the plasma fluctuations are characterized by non-Gaussian PDFs. It has been recognized that the nature of the cross-field transport through the SOL is dominated by turbulence with a significant ballistic or non-local component where a diffusive description is improper \cite{Naulin}. Moreover, the scaling of the confinement time $\tau \propto L^{\alpha}$ with $\alpha < 2$ \cite{kaye} is typical in low-confinement mode discharges, instead of the diffusion induced result $\tau \propto L^2$, where $L$ is the system size. There is a considerable amount of experimental evidence \cite{cardozo1995, gentle1995, callen1997, mantica1999, vanmilligen2002, BalescuBook} and recent numerical gyrokinetic \cite{pradalier, sanchez2008} and fluid \cite{negrete2005} simulations that plasma turbulence in tokamaks is highly non-local.

In addition, intermittent turbulence is characterized by patchy spatial structure that is bursty in time. The PDFs of these intermittent events shows unimodal structure with "elevated" tails that deviates from a Gaussian prediction. The understanding of these events are at best limited \cite{Zweben, bramwell2009, carreras1996, carreras1999, anderson1, anderson2}. Moreover, the high possibility of confinement degradation by intermittency strongly calls for a predictive theory. 

A prominent candidate for explaining the suggestive non-local features of plasma turbulence is the inclusion of a fractional velocity derivative in the Fokker-Planck (FP) equation leading to an inherently non-local description as well as giving rise to non-Gaussian PDFs of e.g. densities and heat flux. The non-locality is introduced through the integral description of the fractional derivative \cite{zaslavsky, sanchez, negrete} and the non-Maxwellian distribution function drives the observed PDFs of densities and heat flux far from Gaussian. 

The aim of this study is to elucidate the effects of a non-Maxwellian distribution function induced by the fractional velocity derivative in the Fokker-Planck equation. Some previous papers on plasma transport have used models including a fractional derivative where the fractional derivative is introduced on phenomenological premises \cite{sanchez, negrete}. In the present work we introduce the Levy statistics into the Langevin equation thus yielding a fractional FP description. This approach is similar to that of Ref. \cite{sanchez2006} resulting in a phenomenological description of the non-local effects in plasma turbulence. Using fractional generalizations of the Liouville equation, kinetic descriptions have been developed previously \cite{zaslovsky,tarasov}. It has been shown that the chaotic dynamics can be described by using the FP equation with coordinate fractional derivatives as a possible tool for the description of anomalous diffusion \cite{zaslovsky2}. Much work has been devoted on investigation of the Langevin equation with Levy white noise, see References \cite{West, Jeperson, Fogedby, Vlad}, or related fractional FP equation \cite{West}. Furthermore, fractional derivatives have been introduced into the FP framework in a similar manner as the present work \cite{chechkin2000, chechkin2002} but a study including drift waves is still called for. To this end we quantify the effects of the fractional derivative in the FP equation in terms of a modified dispersion relation for density gradient driven linear plasma drift waves where we have considered a case with constant external magnetic field and a shear-less slab geometry. In order to calculate an equilibrium PDF we use a model based on the motion of a charged Levy particle in a constant external magnetic field obeying non-Gaussian, Levy statistics. This assumption is the natural generalization of the classical example of the motion of a charged Brownian particle with the usual Gaussian statistics \cite{chandrasekhar}. The fractional derivative is represented with the Fourier transform containing a fractional exponent. We find a relation for the deviation from Maxwellian distribution described by $\epsilon$ through the quasi-neutrality condition and the characteristics of the plasma drift wave are fundamentally changed, i.e. the values of the growth-rate $\gamma$ and real frequency $\omega$ are significantly altered. A deviation from the Maxwellian distribution function alters the dispersion relation for the density gradient drift waves such that the growth rates are substantially increased and thereby may cause enhanced levels of transport. 

The paper is organized as follows. In Sec. 2 the mathematical framework of the fractional FP equation (FFPE) is introduced. In Sec. 3 a dispersion relation for the density gradient driven drift modes using the FFPE are derived. In Sec 4 the deviations from a Maxwellian distribution function are investigated and the dispersion relation is solved in Sec. 5. We conclude the paper with a results and discussion in Sec. 6.

\section{Fractional Fokker-Planck Equation}
Following the theory of Brownian motion we write an equation of motion for a colloidal particle in a background medium as a Langevin equation of the following form \cite{chandrasekhar}
\begin{eqnarray}\label{eq:1.1}
\frac{d\mathbf{v}}{dt}=-\nu\mathbf{v}+A(t)
\end{eqnarray} 
Here, we assumed that the influence of the background medium can be split into a dynamical friction, $-\nu\mathbf{v}$, and a fluctuating part, $A(t)$ which is a Gaussian white noise. The Gaussian white noise assumption is usually imposed in order to obtain a Maxwellian velocity distribution describing the equilibrium of the Brownian particle. This connection is due to the relation between the Gaussian central limit theorem and classical Boltzmann-Gibbs statistics \cite{Khintchine}. However, the Gaussian central limit theorem is not unique and a generalization of the Gaussian central limit theorem to the case of summation of independent identically distributed random variables described by long tailed distributions is performed by L\'{e}vy \cite{levy}, and Khintchine \cite{Khintchine}. In this case the L\'{e}vy distributions replace the Gaussian in a generalized central limit theorem. 

The simplest case of generalized Brownian motion considered by West and Seshadri \cite{seshadri} is to assume for fluctuation part, $A(t)$, in Equation (\ref{eq:1.1}) to be a white L\'{e}vy noise. Following the approach used by Barkai \cite{barkai} we find the Fractional Fokker-Planck Equation (FFPE) with fractional velocity derivatives for shear-less slab geometry in the presence of a constant external force as
\begin{eqnarray}\label{eq:1.2}
\frac{\partial F_{s}}{\partial t}+\mathbf{v}\frac{\partial F_{s}}{\partial \mathbf{r}}+\frac{\mathbf{F}}{m_{s}}\frac{\partial F_{s}}{\partial \mathbf{v}}=\nu\frac{\partial }{\partial \mathbf{v}}(\mathbf{v}F_{s})+D\frac{\partial^{\alpha} F_{s}}{\partial |\mathbf{v}|^{\alpha}},
\end{eqnarray}
where $s(=e,i)$ represents the particle species and $0\le\alpha\le 2$. Here, the term $\frac{\partial^{\alpha} F_{s}}{\partial |\mathbf{v}|^{\alpha}}$ is the fractional Riesz derivative. The fractional differentiation may be represented through singular integrals or by its Fourier transform as we will see later in Equation (\ref{eq:1.4}). Note that the connection to the integral representation indicates that the model is inherently non-local in velocity space. The diffusion coefficient, $D$, is related to the damping term $\nu$, according to a generalized Einstein relation \cite{barkai} 
\begin{eqnarray}\label{eq:1.3}
D=\frac{2^{\alpha-1}T_{\alpha}\nu}{\Gamma(1+\alpha)m_{s}^{\alpha-1}}.
\end{eqnarray}
Here, $T_{\alpha}$ is a generalized temperature, and taking force $\mathbf{F}$ to represent the Lorentz force (due to a constant magnetic field and a zero-averaged electric field) acting on the particles of species $s$ with mass $m_{s}$ and $\Gamma(1+\alpha)$ is the Euler gamma function. We find the solution by using the Fourier representation of equation (\ref{eq:1.2}) above as
\begin{eqnarray}\label{eq:1.4}
\frac{\partial \mathcal{F}_{s}}{\partial t}+(-\mathbf{k}+\Omega_{s}(\mathbf{k}^{v}\times \hat{b})+\nu\mathbf{k}^{v})\frac{\partial\mathcal{F}_{s}}{\partial \mathbf{k}^{v}}=-D|\mathbf{k}^{v}|^{\alpha} \mathcal{F}_{s},
\end{eqnarray}  
where $\Omega_{s}=e_{s}B/m_{s}c$ is the Larmor frequency of species $s$, $\hat{b}=\mathbf{B}/B$ is the unit vector in the direction of magnetic field and $\mathcal{F}_{s}$ is the characteristic function
\begin{eqnarray}\label{eq:1.5}
\mathcal{F}_{s}(\mathbf{k},\mathbf{k}^{v};t)=\int\int d\mathbf{r}\;d\mathbf{v}\exp(i\mathbf{k}\cdot\mathbf{r}+i\mathbf{k}^{v}\cdot\mathbf{v})F_{s}(\mathbf{r},\mathbf{v};t),
\end{eqnarray}
where we have denoted the wave-vector by $\mathbf{k}$ and the corresponding wave vector for the velocity as $\mathbf{k}^v$. We can rewrite the kinetic equation by identification of time derivatives of the wave vectors as
\begin{eqnarray}\label{eq:1.6}
\frac{d\mathcal{F}_{s}}{dt}=\frac{\partial \mathcal{F}_{s}}{\partial t}+\frac{d\mathbf{k}^{v}}{dt}\frac{\partial\mathcal{F}_{s}}{\partial \mathbf{k}^{v}}+\frac{d\mathbf{k}}{dt}\frac{\partial\mathcal{F}_{s}}{\partial \mathbf{k}}=0.
\end{eqnarray}
We use the method of characteristics on the Equation (\ref{eq:1.4}) and (\ref{eq:1.6}) whereby we find that the characteristics are
\begin{eqnarray}\label{eq:1.7}
\frac{\partial \mathcal{F}_{s}}{\partial t}=-D|\mathbf{k}^{v}|^{\alpha} \mathcal{F}_{s},\\
\frac{d\mathbf{k}^{v}}{dt}=-\mathbf{k}+\Omega_{s}(\mathbf{k}^{v}\times \hat{b})+\nu\mathbf{k}^{v},\\
\frac{d\mathbf{k}}{dt}=0.
\end{eqnarray}
Following the method used in Ref. \cite{chechkin2002,chechkin2000} the solution corresponding to the homogenous and steady state system in Fourier space is 
\begin{eqnarray}\label{eq:2.14}
\mathcal{F}_{s}(\mathbf{k}^{v}, t)=e^{-\frac{D}{\alpha\nu}(|\mathbf{k}^{v}_{\bot}|^{\alpha}+|\mathbf{k}^{v}_{\parallel}|^{\alpha})}.
\end{eqnarray}
In order to find the solution in real space we compute the inverse Fourier transform of Equation (\ref{eq:2.14}) 
\begin{eqnarray}\label{eq:2.15}
F_{s}(\mathbf{r},\mathbf{v})=C(\mathbf{r})\int \frac{d\mathbf{k}_{\bot}^{v}d\mathbf{k}_{\parallel}^{v}}{(2\pi)^{3/2}}e^{-i(\mathbf{k}_{\bot}^{v}\mathbf{v}_{\bot}+\mathbf{k}_{\parallel}^{v}\mathbf{v}_{\parallel})}e^{-\frac{D}{\alpha\nu}(|\mathbf{k}^{v}_{\bot}|^{\alpha}+|\mathbf{k}^{v}_{\parallel}|^{\alpha})}.
\end{eqnarray} 
We define a new variable $\mathcal{D}=\frac{D}{\nu}$ where coefficient $D$ is given by the expression in Equation (\ref{eq:1.3}). $C(\mathbf{r})$ is a normalization factor which remains to be defined. Taking the inverse Fourier transform of the Equation (\ref{eq:2.15}) for $\alpha=2$ we get
\begin{eqnarray}\label{eq:2.16}
F_{s}(\mathbf{r},\mathbf{v})=\frac{C(\mathbf{r})}{\mathcal{D}}e^{-(\frac{v_{\bot}^2+v_{\parallel}^2}{4\mathcal{D}})}.
\end{eqnarray}
The unknown normalization factor $C$ can be determined by comparing the integrals of the Maxwellian distribution and our distribution. In comparison the Maxwellian distribution is defined as
\begin{eqnarray}\label{eq:2.17}
F^{M}_{s}(\mathbf{r},\mathbf{v})=\frac{n_{s}(\mathbf{r})}{(\sqrt{\pi}V_{T,s}(\mathbf{r}))^3}e^{-(v_{\bot}^2+v_{\parallel}^2)/V_{T,s}^2(\mathbf{r})},
\end{eqnarray}
where $V_{T,s}(\mathbf{r})=\sqrt{2T_{s}(\mathbf{r})/m_{s}}$ is the thermal velocity of species $s$. By integrating the Maxwellian distribution over the velocity space we find the density as
\begin{eqnarray}\label{eq:2.18}
\int d\mathbf{v}F^{M}_{s}(\mathbf{r},\mathbf{v})=2\pi\int_{0}^{\infty}v_{\bot}dv_{\bot}\int_{-\infty}^{\infty}dv_{\parallel}\frac{n_{s}(\mathbf{r})}{(\sqrt{\pi}V_{T,s}(\mathbf{r}))^3}e^{-(v_{\bot}^2+v_{\parallel}^2)/V_{T,s}^2(\mathbf{r})}=n_{s}(\mathbf{r}),
\end{eqnarray}
whereas performing the same integration of the expression in Equation(\ref{eq:2.16}) we obtain
\begin{eqnarray}\label{eq:2.19}
\int d\mathbf{v}F_{s}(\mathbf{r},\mathbf{v})=2\pi\int_{0}^{\infty}v_{\bot}dv_{\bot}\int_{-\infty}^{\infty}dv_{\parallel}\frac{C(\mathbf{r})}{ \mathcal{D}}e^{-(\frac{v_{\bot}^2+v_{\parallel}^2}{4\mathcal{D}})}=2\pi^{3/2}\sqrt{2\mathcal{D}} C(\mathbf{r}). 
\end{eqnarray}
We can now compare the two results obtained in Equations (\ref{eq:2.18}) - (\ref{eq:2.19}) and we find the following relation
\begin{eqnarray}\label{eq:1.21}
C(\mathbf{r})=\frac{n_{s}(\mathbf{r})}{2\pi^{3/2}\sqrt{2\mathcal{D}}}. 
\end{eqnarray}
The distribution function can now be determined by replacing this expression into Equation (\ref{eq:2.16}) for $C(\mathbf{r})$ yielding
\begin{eqnarray}\label{eq:1.22}
F_{s}(\mathbf{r},\mathbf{v})=\frac{n_{s}(\mathbf{r})}{2\pi^{3/2}\mathcal{D}\sqrt{2\mathcal{D}}} e^{-(\frac{v_{\bot}^2+v_{\parallel}^2}{4\mathcal{D}})}.
\end{eqnarray}
We can easily recover the Maxwellian distribution in Equation (\ref{eq:2.17}) by setting $\alpha=2$ in the definition for $D$ in Equation (\ref{eq:1.3}) and using that $\Gamma(3)=2$. Note that for a general $\alpha$, the equilibrium distribution is as follows
\begin{eqnarray}\label{eq:2.24.1}
F_{s}(\mathbf{r},\mathbf{v})=\frac{n_{s}(\mathbf{r})}{2\pi^{3/2}\sqrt{2\mathcal{D}}} \int \frac{d\mathbf{k}_{\bot}^{v}d\mathbf{k}_{\parallel}^{v}}{(2\pi)^{3/2}}e^{-i(\mathbf{k}_{\bot}^{v}\mathbf{v}_{\bot}+\mathbf{k}_{\parallel}^{v}\mathbf{v}_{\parallel})}e^{-\frac{\mathcal{D}}{\alpha}(|\mathbf{k}^{v}_{\bot}|^{\alpha}+|\mathbf{k}^{v}_{\parallel}|^{\alpha})},
\end{eqnarray}
where
\begin{eqnarray}\label{eq:2.25.1}
\mathcal{D}=\frac{V_{T,s}^{\alpha}}{\Gamma(1+\alpha)},
\end{eqnarray}
and we have introduced a generalized thermal velocity as 
\begin{eqnarray}\label{eq:2.25.2}
V_{T,s}^{\alpha}=\frac{2^{\alpha-1}T_{\alpha}}{m_{s}^{\alpha-1}}.
\end{eqnarray}
The generalized equilibrium distribution including the effects of the fractional velocity derivative in Equation (\ref{eq:2.24.1}) becomes
\begin{eqnarray}\label{eq:2.24}
F_{s}(\mathbf{r},\mathbf{v})=\frac{n_{s}(\mathbf{r})}{2\pi^{3/2}(\Gamma(1+\alpha))^{-1/2}\sqrt{2V_{T,s}^{\alpha}}} 
\int \frac{d\mathbf{k}_{\bot}^{v}d\mathbf{k}_{\parallel}^{v}}{(2\pi)^{3/2}}e^{-i(\mathbf{k}_{\bot}^{v}\mathbf{v}_{\bot}+\mathbf{k}_{\parallel}^{v}\mathbf{v}_{\parallel})}e^{-\frac{V_{T,s}^{\alpha}}{\Gamma(1+\alpha)\alpha}(|\mathbf{k}^{v}_{\bot}|^{\alpha}+|\mathbf{k}^{v}_{\parallel}|^{\alpha})}.
\end{eqnarray}
We will now determine the dispersion relation for density gradient driven drift waves including the effects of the fractional velocity differential operator. 

%%%%%%%%%%%%%%%%%%%%%%%%%%%%%%%%%%%%%%%
\section{The dispersion relation}
%%%%%%%%%%%%%%%%%%%%%%%%%%%%%%%%%%%%%%%
In order to quantify the non-local effects on drift waves induced by the fractional differential operator we will determine the dispersion relation for density gradient driven drift modes. We start by formulating the linearized gyro-kinetic theory where the particle distribution function, averaged over gyro-phase is of the form (see Ref. \cite{Balescu1991})
\begin{eqnarray}\label{eq:2.26}
f_{s}(\mathbf{r},\mathbf{v})=F_{s}(\mathbf{r},\mathbf{v})+(2\pi)^{-4}\times\int\int d\mathbf{k}\;d\omega\exp(i\mathbf{k}\cdot\mathbf{r}-i\omega t)\delta f^{s}_{\mathbf{k},\omega}(\mathbf{v}).
\end{eqnarray} 
We assume that the turbulence is purely electrostatic and neglect magnetic field fluctuations $(\delta \mathbf{B}=0)$. For small deviations from the local equilibrium we find the linearized gyro-kinetic equation of the form
\begin{eqnarray}\label{eq:2.27}
(-\omega+k_{\parallel}v_{\parallel})\delta f^{s}_{\mathbf{k},\omega}(v_{\parallel},v_{\bot})+(\omega-\omega_{*s})\frac{e_{s}}{T_{s}}J_{0}(|\Omega_{s}|^{-1}k_{\bot}v_{\bot})F_{s}(x,\mathbf{v})\delta \phi_{\mathbf{k},\omega}=0,
\end{eqnarray} 
where $\omega_{*s}=\frac{cT_{s}}{e_{s}B}k_{y}\cdot\frac{d\; ln\;n(x)}{d x}$ is the drift wave frequency of species $s$, and we assumed that the space dependence of $F_{s}$ is only in the $x$ direction perpendicular to the magnetic field as well as for the density gradient. In the equation above, $J_{0}$ is the Bessel function of order zero, $v_{\parallel}$ is the parallel velocity, $v_{\bot}\equiv (v_{x}^{2}+v_{y}^{2})^{1/2}$ is the perpendicular velocity and hence we write the total speed as $v=(v_{\bot}^{2}+v_{\parallel}^{2})^{1/2}$. Inserting the expression for $F_{s}$ from the Equation (\ref{eq:2.24}) in Equation (\ref{eq:2.27}) and rearranging the terms we find the perturbed distribution $\delta f_{\mathbf{k},\omega}$ as
\begin{eqnarray}\label{eq:2.28}
\delta f^{s}_{\mathbf{k},\omega}(v_{\parallel},v_{\bot})=
-\frac{e_{s}}{T_{s}}[\frac{\omega-\omega_{*s}}{k_{\parallel}v_{\parallel}-\omega}]J_{0}(|\Omega_{s}|^{-1}k_{\bot}v_{\bot})\delta \phi_{\mathbf{k},\omega}
\frac{n_{s}(\mathbf{r})}{2\pi^{3/2}(\Gamma(1+\alpha))^{-1/2}\sqrt{2V_{T,s}^{\alpha}}} \times\nonumber\\
\int \frac{d\mathbf{k}_{\bot}^{v}d\mathbf{k}_{\parallel}^{v}}{(2\pi)^{3/2}}e^{-i(\mathbf{k}_{\bot}^{v}\mathbf{v}_{\bot}+\mathbf{k}_{\parallel}^{v}\mathbf{v}_{\parallel})}e^{-\frac{V_{T,s}^{\alpha}}{\Gamma(1+\alpha)\alpha}(|\mathbf{k}^{v}_{\bot}|^{\alpha}+|\mathbf{k}^{v}_{\parallel}|^{\alpha})}.
\end{eqnarray}
Here, the wave vector perpendicular to magnetic field is $k_{\bot}=(k^2_{x}+k^2_{y})^{1/2}$. The gyro-kinetic Equation (\ref{eq:2.28}) is complemented with Poisson equation for the electric potential. For fluctuations with wave vectors much smaller than the Debye wave vector, the Poisson equation becomes the quasi-neutrality condition
\begin{eqnarray}\label{eq:2.29}
\sum_{s} e_{s}\delta n^{s}_{\mathbf{k},\omega}=0,
\end{eqnarray}
where the density fluctuation is related to the distribution function through
\begin{eqnarray}\label{eq:2.31}
\delta n^{s}_{\mathbf{k},\omega}=-\frac{e_{s}}{T_{s}}n_{s}\delta\phi_{\mathbf{k},\omega} + \int d\mathbf{v}
J_{0}(|\Omega_{s}|^{-1}k_{\bot}v_{\bot})\delta f^{s}_{\mathbf{k},\omega}(v_{\parallel},v_{\bot}).
\end{eqnarray}
In the above equation we have separated the adiabatic response (first term on the right hand side) from the non-adiabatic response (second term on the right hand side). We have to keep in mind that the density $n_{s}$ coming from the $F_{s}(x,\mathbf{v})$ in the adiabatic response is also given by Equation (\ref{eq:2.24}) and for a general $0\le\alpha\le2$ the adiabatic response can be different than that calculated by Maxwellian distribution of Equation (\ref{eq:2.17}). Using the quasi-neutrality condition (\ref{eq:2.29}) we find the dispersion equation which determines the eigenfrequencies as a function of the wave vector, $\omega=\omega(\mathbf{k})=\omega_{r}(\mathbf{k})+i\gamma(\mathbf{k})$. In the simplest case we consider a plasma consisting of electrons and a single species of singly charged ions with the equal temperatures. For the density fluctuation therefore we have
\begin{eqnarray}\label{eq:2.32}
\delta n^{s}_{\mathbf{k},\omega}=-n_{s}(\mathbf{r})\frac{e_{s}}{T_{s}}\delta\phi_{\mathbf{k},\omega}[M^{ad,s}+M^{s}_{\mathbf{k},\omega}].
\end{eqnarray}
Therefore, the dispersion equation as in the Ref. \cite{Balescu1991} is
\begin{eqnarray}\label{eq:2.33}
M^{ad,e}+M^{e}_{\mathbf{k},\omega}=-M^{ad,i}-M^{i}_{\mathbf{k},\omega},
\end{eqnarray}
where
\begin{eqnarray}\label{eq:2.34}
M^{ad,s}=\int d\mathbf{v}
\frac{1}{2\pi^{3/2}(\Gamma(1+\alpha))^{-1/2}\sqrt{2V_{T,s}^{\alpha}}} 
\int \frac{d\mathbf{k}_{\bot}^{v}d\mathbf{k}_{\parallel}^{v}}{(2\pi)^{3/2}}e^{-i(\mathbf{k}_{\bot}^{v}\mathbf{v}_{\bot}+\mathbf{k}_{\parallel}^{v}\mathbf{v}_{\parallel})}e^{-\frac{V_{T,s}^{\alpha}}{\Gamma(1+\alpha)\alpha}(|\mathbf{k}^{v}_{\bot}|^{\alpha}+|\mathbf{k}^{v}_{\parallel}|^{\alpha})},
\end{eqnarray}
gives the adiabatic contribution, and
\begin{eqnarray}\label{eq:2.35}
M^{s}_{\mathbf{k},\omega}=\int d\mathbf{v}[\frac{\omega-\omega_{*s}}{k_{\parallel}v_{\parallel}-\omega}]J_{0}(b_{s}v_{\bot}/V_{Ts})\times\nonumber\\
\frac{1}{2\pi^{3/2}(\Gamma(1+\alpha))^{-1/2}\sqrt{2V_{T,s}^{\alpha}}}
\int \frac{d\mathbf{k}_{\bot}^{v}d\mathbf{k}_{\parallel}^{v}}{(2\pi)^{3/2}}e^{-i(\mathbf{k}_{\bot}^{v}\mathbf{v}_{\bot}+\mathbf{k}_{\parallel}^{v}\mathbf{v}_{\parallel})}e^{-\frac{V_{T,s}^{\alpha}}{\Gamma(1+\alpha)\alpha}(|\mathbf{k}^{v}_{\bot}|^{\alpha}+|\mathbf{k}^{v}_{\parallel}|^{\alpha})},
\end{eqnarray}
gives the non-adiabatic contribution. Here, $b_{s}=k_{\bot}V_{T,s}/\Omega_{s}$. If we take $\alpha=2$ in the Equation(\ref{eq:2.33}) we recover the dispersion equation for a Maxwellian distribution as in the Ref. \cite{Balescu1991}.

\subsection{Adiabatic response}
First, we may analyze the contribution from the adiabatic parts of the dispersion relation only by ignoring all fluctuations, yielding \begin{eqnarray}\label{eq:2.37}
|M^{ad,e}|=|M^{ad,i}|.
\end{eqnarray} 
In addition, utilizing the quasi-neutrality condition while neglecting the density gradient in the system we have $n_{i}=n_{e}$, therefore $\alpha_{e}$ and $\alpha_{i}$ becomes connected through Equation (\ref{eq:2.37}). This indicates that the deviation from a Maxwellian distribution described by $\alpha$ for electrons and ions becomes dependent on each other. We will get back to this relation in later sections.

%%%%%%%%%%%%%%%%%%%%%%%%%%%%%%%%%%%%%%%%%
\section{Deviations from a Maxwellian distribution function}
%%%%%%%%%%%%%%%%%%%%%%%%%%%%%%%%%%%%%%%%%
We will now turn our attention to the problem of solving the dispersion relation described by Equation (\ref{eq:2.33}). In order to solve this dispersion equation we use the method proposed in Ref. \cite{Balescu1991} with the difference that here we have to perform additional integrations over $\mathbf{k}^{v}$. We have
\begin{eqnarray}\label{eq:3.1}
M^{s}_{\mathbf{k},\omega}=\frac{\omega-\omega_{*,s}}{|k_{\parallel}|V_{T,s}}Z(\xi_{s})\Gamma(b_{s}),
\end{eqnarray}
where the plasma dispersion function is
\begin{eqnarray}\label{eq:3.2}
Z(\xi_{s})=\frac{V_{T,s}}{\sqrt{\pi}}Lim_{\sigma\rightarrow 0}\int_{-\infty}^{\infty}du[\frac{\Phi(v_{\parallel})}{u-\xi_{s}-i\sigma}],
\end{eqnarray}
with $u=v_{\parallel}/V_{Ts}$,  $\xi_{s}=\omega/(|k_{\parallel}|V_{Ts})$ and the function $\Phi(v_{\parallel})$ is
\begin{eqnarray}\label{eq:3.3}
\Phi(v_{\parallel})=\frac{1}{\sqrt{2(\Gamma(1+\alpha))^{-1/2}\sqrt{2V_{T,s}^{\alpha}}}}
\int \frac{d\mathbf{k}_{\parallel}^{v}}{(2\pi)^{1/2}}e^{-i\mathbf{k}_{\parallel}^{v}\mathbf{v}_{\parallel}}e^{-\frac{V_{T,s}^{\alpha}}{\Gamma(1+\alpha)\alpha}(|\mathbf{k}^{v}_{\parallel}|^{\alpha})}.
\end{eqnarray}
The integral over $v_{\bot}$ can be written in a general way as
\begin{eqnarray}\label{eq:3.4}
\Gamma(b_{s})=2V_{T,s}^2\int_{0}^{\infty}dw w \Psi_{s}(b_{s}w)\Phi(v_{\bot}),
\end{eqnarray}
where $w=v_{\bot}/V_{Ts}$, $\Psi_{s}=J_{0}^2(b_{s}v_{\bot}/V_{Ts})$ and,
\begin{eqnarray}\label{eq:3.5}
\Phi(v_{\bot})=\frac{1}{\sqrt{2(\Gamma(1+\alpha))^{-1/2}\sqrt{2V_{T,s}^{\alpha}}}} 
\int \frac{d\mathbf{k}_{\bot}^{v}}{(2\pi)}e^{-i\mathbf{k}_{\bot}^{v}\mathbf{v}_{\bot}}e^{-\frac{V_{T,s}^{\alpha}}{\Gamma(1+\alpha)\alpha}(|\mathbf{k}^{v}_{\bot}|^{\alpha})}.
\end{eqnarray}
The analytical solutions for integrals over $\mathbf{k}^{v}$ with an arbitrary $\alpha$ in the Equations (\ref{eq:3.3}) and (\ref{eq:3.5}) requires rather tedious calculations. Instead we consider an infinitesimal deviation of the form $\alpha=2-\epsilon$, where $0\le\epsilon\ll 2$ and expand the terms depending on $\alpha$ in the Equations (\ref{eq:3.3}) and (\ref{eq:3.5}) around $\epsilon=0$ as follows
\begin{eqnarray}\label{eq:3.6}
\frac{1}{\sqrt{(\Gamma(1+\alpha))^{-1/2}\sqrt{V_{T,s}^{\alpha}}}} e^{-\frac{V_{T,s}^{\alpha}}{\Gamma(1+\alpha)\alpha}(|k^{v}|^{\alpha})}=
\frac{2^{1/4}e^{-\frac{1}{4}V_{T,s}^{2}|k^{v}|^{2}}}{\sqrt{V_{T,s}}}+\epsilon\Lambda(k^{v})+\mathcal{O}[\epsilon^2],
\end{eqnarray}
where
\begin{eqnarray}\label{eq:3.7}
\Lambda(k^{v})=\frac{e^{-\frac{1}{4}V_{T,s}^{2}|k^{v}|^{2}}}{2^{11/4}\sqrt{V_{T,s}}}\{
-3+2{\gamma_E}-4V_{T,s}^2|k^{v}|^2+2 {\gamma_E} V_{T,s}^2|k^{v}|^2\nonumber\\
+2{\log}[V_{T,s}]+2V_{T,s}^2 {\log}[V_{T,s}]|k^{v}|^2+2V_{T,s}^2|k^{v}|^2 {\log}[|k^{v}|]\}.
\end{eqnarray}
Here, we have used the Euler-Mascheroni constant $\gamma_E = 0.57721$. The first term in Equation (\ref{eq:3.6}) will produce
\begin{eqnarray}\label{eq:3.3.1}
\Phi(u)=\frac{e^{-u^2}}{V_{T,s}^{3/2}},\;\;\;\;\;\;\; \mbox{and} \;\;\;\;\;\;\;\;\Phi(w)=\frac{e^{-w^2}}{V_{T,s}^{3/2}}
\end{eqnarray}
which give the Maxwellian adiabatic response
\begin{eqnarray}\label{eq:3.1.1}
M^{ad,s}=1.
\end{eqnarray}
By using the expansion defined by the expression (\ref{eq:3.6}) in Equations (\ref{eq:2.34}) and (\ref{eq:2.34}), the adiabatic and non-adiabatic part of the dispersion relation $M^{ad,s}$ and $M^{s}_{\mathbf{k},\omega}$ are as follows
\begin{eqnarray}\label{eq:3.8}
M^{ad,s}=1+(2\pi\int_{-\infty}^{\infty}dv_{\parallel}\int_{0}^{\infty}dv_{\bot}v_{\bot}\times\nonumber\\
\frac{1}{2\sqrt{2}\pi^{3/2}} \int \frac{d\mathbf{k}_{\bot}^{v}d\mathbf{k}_{\parallel}^{v}}{(2\pi)^{3/2}}e^{-i(\mathbf{k}_{\bot}^{v}\mathbf{v}_{\bot}+\mathbf{k}_{\parallel}^{v}\mathbf{v}_{\parallel})}\Lambda(k_{\bot}^{v})\Lambda(k_{\parallel}^{v}))\epsilon+\mathcal{O}[\epsilon]^2=1+\epsilon W^{ad,s}.\nonumber\\
\end{eqnarray}
and
\begin{eqnarray}\label{eq:3.9}
M^{s}_{\mathbf{k},\omega}=2\pi\int_{-\infty}^{\infty}dv_{\parallel}\int_{0}^{\infty}dv_{\bot}v_{\bot}[\frac{\omega-\omega_{*s}}{k_{\parallel}v_{\parallel}-\omega}]\Psi_{s}(b_{s}v_{\bot}/V_{Ts})\times\nonumber\\
\frac{1}{(\sqrt{\pi}V_{T,s}(\mathbf{r}))^3}e^{-(v_{\bot}^2+v_{\parallel}^2)/V_{T,s}^2(\mathbf{r})}+\nonumber\\
(2\pi\int_{-\infty}^{\infty}dv_{\parallel}\int_{0}^{\infty}dv_{\bot}v_{\bot}[\frac{\omega-\omega_{*s}}{k_{\parallel}v_{\parallel}-\omega}]\Psi_{s}(b_{s}v_{\bot}/V_{Ts})\times\nonumber\\
\frac{1}{2\sqrt{2}\pi^{3/2}} \int \frac{d\mathbf{k}_{\bot}^{v}d\mathbf{k}_{\parallel}^{v}}{(2\pi)^{3/2}}e^{-i(\mathbf{k}_{\bot}^{v}\mathbf{v}_{\bot}+\mathbf{k}_{\parallel}^{v}\mathbf{v}_{\parallel})}\Lambda(k_{\bot}^{v})\Lambda(k_{\parallel}^{v}))\epsilon+\mathcal{O}[\epsilon]^2=N^{s}_{\mathbf{k},\omega}+\epsilon W^{s}_{\mathbf{k},\omega}.\nonumber\\
\end{eqnarray}
Inserting these relations we may rewrite the dispersion relation (\ref{eq:2.33}) in the form
\begin{eqnarray}\label{eq:3.10}
(1+N^{e}_{\mathbf{k},\omega})+\epsilon (W^{ad,e}+W^{e}_{\mathbf{k},\omega})=-(1+N^{i}_{\mathbf{k},\omega})-\epsilon (W^{ad,i}+W^{i}_{\mathbf{k},\omega}).
\end{eqnarray}
The first terms on the right and left hand sides generate the usual contributions to the dispersion equation as in Ref. \cite{Balescu1991} and the terms proportional to $\epsilon$ generate the non-Maxwellian contributions where we have
\begin{eqnarray}\label{eq:3.11}
N^{s}_{\mathbf{k},\omega}=\frac{\omega-\omega_{*,s}}{|k_{\parallel}|V_{T,s}}Z(\xi_{s})\Gamma(b_{s}),
\end{eqnarray}
with the usual plasma dispersion function $Z(\xi_{s})$ written as
\begin{eqnarray}\label{eq:3.12}
Z(\xi_{s})=\frac{1}{\sqrt{\pi}}Lim_{\sigma\rightarrow 0}\int_{-\infty}^{\infty}du e^{-u^2}[\frac{1}{u-\xi_{s}-i\sigma}],
\end{eqnarray}
and 
\begin{eqnarray}\label{eq:3.13}
\Gamma(b_{s})=2\int_{0}^{\infty}dw w e^{-w^2}\Psi_{s}(b_{s}w).
\end{eqnarray}
The effects of the fractional velocity derivative can be boiled down to a non-Maxwellian contribution of the form
\begin{eqnarray}\label{eq:3.14}
W^{s}_{\mathbf{k},\omega}=\frac{\omega-\omega_{*,s}}{|k_{\parallel}|V_{T,s}}Z_{\epsilon}(\xi_{s})\Gamma_{\epsilon}(b_{s}),
\end{eqnarray}
where the non-Maxwellian plasma dispersion function is given by
\begin{eqnarray}\label{eq:3.15}
Z_{\epsilon}(\xi_{s})=\frac{V_{T,s}}{\sqrt{\pi}}Lim_{\sigma\rightarrow 0}\int_{-\infty}^{\infty}du[\frac{\Phi(v_{\parallel})}{u-\xi_{s}-i\sigma}],
\end{eqnarray}
with the function $\Phi(v_{\parallel})$ being
\begin{eqnarray}\label{eq:3.16}
\Phi(v_{\parallel})=\frac{1}{2^{3/4}}\int \frac{dk^{v}_{\parallel}}{(2\pi)^{1/2}}\exp(-ik^{v}_{\parallel}v_{\parallel})\Lambda(k_{\parallel}^{v}).
\end{eqnarray}
It is important to note that the deviation from Maxwellian is different for the different species (electrons and ions). In the rest of Sec. 4, we will quantify the deviations. The non-Maxwellian contribution to Equation (\ref{eq:3.4}) is
\begin{eqnarray}\label{eq:3.17}
\Gamma_{\epsilon}(b_{s})=2V_{T,s}^2\int_{0}^{\infty}dw w \Psi_{s}(b_{s}w)\Phi(v_{\bot}),
\end{eqnarray}
where
\begin{eqnarray}\label{eq:3.18}
\Phi(v_{\bot})=\frac{1}{2^{3/4}}\int \frac{dk^{v}_{\bot}}{(2\pi)}\exp(-ik^{v}_{\bot}v_{\bot})\Lambda(k_{\bot}^{v}).
\end{eqnarray}
To extimate the non-Maxwellian contribution we need to determine the inverse Fourier transforms of the Equations (\ref{eq:3.16}) and (\ref{eq:3.18}) resulting in
\begin{eqnarray}\label{eq:3.18.1}
\Phi(z)=\frac{1}{8V_{T,s}^{3/2}}e^{-z^2} \nonumber \\ 
\left\{-4(-2+{\gamma_E})z^2+(-7+4\; {\gamma_E})+ 2 \log[V_{T,s}]+2e^{z^2}{_1 F_1}[\frac{3}{2},\frac{1}{2},-z^2]\right\}
\end{eqnarray}
with $z=\{u,w\}$ and ${_1 F_1}[a;b;z]$ denoting Kummer's confluent hypergeometric function. Therefore we can write
\begin{eqnarray}\label{eq:3.18.2}
W^{ad,s}= \frac{2V_{T,s}^3}{\sqrt{\pi}}  \int_{-\infty}^{\infty}du\int_{0}^{\infty}wdw \Phi(u)\Phi(w).
\end{eqnarray}
By inserting typical values for the plasma parameters from  Ref. \cite{BalescuBook} we find the velocities as $V_{T,e}=5.93\times 10^{9} [cm/s]$ and $V_{T,i}=1.38\times 10^{8} [cm/s]$ and we obtain
\begin{eqnarray}\label{eq:3.18.3}
W^{ad,e}=33.724\;\;\;\;\;\;\;\;\;,W^{ad,i}=23.6591.
\end{eqnarray}
Following the adiabatic condition in Equation (\ref{eq:2.37}) and the expanded dispersion relation in Equation (\ref{eq:3.10}) we obtain the following ratio between the non-Maxwellian contributions
\begin{eqnarray}\label{eq:3.18.4}
\frac{\epsilon_{i}}{\epsilon_{e}}=\frac{W^{ad,e}}{W^{ad,i}}=1.42541.
\end{eqnarray}
This relation means that if there is a deviation of the distribution function from the Maxwellian for plasma electrons, the deviation from the Maxwellian for ions will be $\sim 1.4$ larger.

%%%%%%%%%%%%%%%%%%%%%%%%%%%%%%%%%%%%%%%%%%%%%%%%%
\section{Solutions of the dispersion relation}
%%%%%%%%%%%%%%%%%%%%%%%%%%%%%%%%%%%%%%%%%%%%%%%%%
We will solve the dispersion relation in terms of expansions of the plasma dispersion function by noting that the drift waves are defined in the frequency range $|k_{\parallel}|V_{Ti}\ll\omega \ll |k_{\parallel}|V_{Te}$ in evaluating Equations (\ref{eq:3.12}) and (\ref{eq:3.15}). We define the expansion parameter for electrons in powers of $\xi_{e}=\omega/(|k_{\parallel}|V_{Te})\ll 1$ and for ions we expand it in powers of $\xi_{i}^{-1}=(|k_{\parallel}|V_{Ti})/\omega\ll 1$, respectively. The Maxwellian dispersion function $Z(\xi_{s})$ has the same definition as in Ref. \cite{Balescu1991}
\begin{eqnarray}\label{eq:3.19}
Z(\xi_{e})=\frac{1}{\sqrt{\pi}}Lim_{\sigma\rightarrow 0}\int_{-\infty}^{\infty}du e^{-u^2}[\frac{1}{u-\xi_{e}-i\sigma}]
=-2\xi_{e}+\frac{4\xi_{e}^3}{3}+i\sqrt{\pi}(1-\xi_{e}^2)+\mathcal{O}[\xi_{e}^4],
\end{eqnarray}
whereas the non-Maxwellian plasma dispersion function $Z_{\epsilon}(\xi_{e})$ becomes
\begin{eqnarray}\label{eq:3.20}
Z_{\epsilon}(\xi_{e})=\frac{V_{T,e}}{\sqrt{\pi}}Lim_{\sigma\rightarrow 0}\int_{-\infty}^{\infty}du[\frac{\Phi(u)}{u-\xi_{e}-i\sigma}]
=\frac{V_{T,e}}{\sqrt{\pi}}Lim_{\sigma\rightarrow 0}\nonumber\\
\int_{-\infty}^{\infty}du\Phi(u)[\frac{1}{u-i\sigma}+\frac{\xi_{e}}{(u-i\sigma)^2}+\frac{\xi_{e}^2}{(u-i\sigma)^3}+\frac{\xi_{e}^3}{(u-i\sigma)^4}+\mathcal{O}[\xi_{e}^4]].
\end{eqnarray}
For ions, using the expansion in powers of $\xi_{i}^{-1}$ we can rewrite the above integrals as a function of the expansion parameter as
\begin{eqnarray}\label{eq:3.21}
Z(\xi_{i})=\frac{1}{\sqrt{\pi}}Lim_{\sigma\rightarrow 0}\int_{-\infty}^{\infty}du e^{-u^2}[\frac{1}{u-\xi_{i}-i\sigma}]
=-\xi_{i}^{-1}-\frac{1}{2}\xi_{i}^{-3}+\mathcal{O}[\xi_{i}^{-5}],
\end{eqnarray}
and the non-Maxwellian $Z_{\epsilon}(\xi_{i})$ becomes
\begin{eqnarray}\label{eq:3.22}
Z_{\epsilon}(\xi_{i})=\frac{V_{T,i}}{\sqrt{\pi}}Lim_{\sigma\rightarrow 0}\int_{-\infty}^{\infty}du[\frac{\Phi(u)}{u-\xi_{i}-i\sigma}]
=\frac{V_{T,i}}{\sqrt{\pi}}Lim_{\sigma\rightarrow 0}\nonumber\\
\int_{-\infty}^{\infty}du\Phi(u)[\frac{1}{(-\xi_{i}-i\sigma)}-\frac{u}{(\xi_{i}+i\sigma)^2}+\frac{u^2}{(-\xi_{i}-i\sigma)^3}-\frac{u^3}{(\xi_{i}+i\sigma)^4}+\mathcal{O}[\xi_{i}^{-5}]].
\end{eqnarray}
We can now evaluate he Maxwellian integrals of the forms $\Gamma(b_{e})$ and $\Gamma(b_{i})$ assuming $\Psi_{e}=1$, $\Psi_{i}=J_{0}^2(b_{i}v_{\bot}/V_{Ti})$ we get
\begin{eqnarray}\label{eq:3.23}
\Gamma(b_{e})=2\int_{0}^{\infty}dw w e^{-w^2}=1,
\end{eqnarray}
and
\begin{eqnarray}\label{eq:3.24}
\Gamma(b_{i})=2\int_{0}^{\infty}dw w e^{-w^2}\Psi_{e}(b_{i}w)=e^{-b_{i}/2}\mathcal{I}_{0}(b_{i}),
\end{eqnarray}
where $\mathcal{I}_{0}$ denotes modified Bessel function of the zeroth order. The final result will be found after evaluating the non-Maxwellian $\Gamma_{\epsilon}(b_{e})$ and $\Gamma_{\epsilon}(b_{i})$ are given as
\begin{eqnarray}\label{eq:3.25}
\Gamma_{\epsilon}(b_{e})=2V_{T,e}^2\int_{0}^{\infty}dw w \Phi(w)=4.8\times 10^{5},
\end{eqnarray}
and
\begin{eqnarray}\label{eq:3.26}
\Gamma_{\epsilon}(b_{i})=2V_{T,i}^2\int_{0}^{\infty}dw w \Psi(b_{i}w)\Phi(w)=6.1\times 10^{4},
\end{eqnarray}
where we have used $V_{T,e}=5.93\;10^{9} [cm/s]$, $V_{T,i}=1.38\;10^{8} [cm/s]$ and $b_{i}=0.1$. Finally we can summarize different terms in the dispersion relation (\ref{eq:3.10}) as
\begin{eqnarray}\label{eq:3.27}
N^{e}_{\mathbf{k},\omega}=(\xi_{e}-\bar{\omega}_{*,e})(-2\xi_{e}+\frac{4\xi_{e}^3}{3}+i\sqrt{\pi}(1-\xi_{e}^2)),\nonumber\\
N^{i}_{\mathbf{k},\omega}=(\xi_{i}-\bar{\omega}_{*,i})(-\xi_{i}^{-1}-\frac{1}{2}\xi_{i}^{-3})e^{-b_{i}/2}\mathcal{I}_{0}(b_{i}),\nonumber\\
W^{e}_{\mathbf{k},\omega}=(\xi_{e}-\bar{\omega}_{*,e})Z_{\epsilon}(\xi_{e})\Gamma_{\epsilon}(b_{e}),\nonumber\\
W^{i}_{\mathbf{k},\omega}=(\xi_{i}-\bar{\omega}_{*,i})Z_{\epsilon}(\xi_{i})\Gamma_{\epsilon}(b_{i}),
\end{eqnarray}
where $\bar{\omega}_{*,s}=\omega_{*,s}/|k_{\parallel}|V_{T,s}$.
Note that the non-Maxwellian contributions in Equations (\ref{eq:3.20}), (\ref{eq:3.22}), (\ref{eq:3.25}) and (\ref{eq:3.26}) have been calculated numerically. By utilizing the found values of the integrals above we rewrite the dispersion relation (\ref{eq:3.10}) as follows
\begin{eqnarray}\label{eq:3.28}
(1+\epsilon_{e} W^{ad,e})+(\xi_{e}-\bar{\omega}_{*,e})\{-2\xi_{e}+\frac{4\xi_{e}^3}{3}+i\sqrt{\pi}(1-\xi_{e}^2)+\epsilon_{e} Z_{\epsilon}(\xi_{e}) \Gamma_{\epsilon}(b_{e})\}=\nonumber\\
-(1+\epsilon_{i} W^{ad,i})-(\xi_{i}-\bar{\omega}_{*,i})\{(-\xi_{i}^{-1}-\frac{1}{2}\xi_{i}^{-3})e^{-b_{i}/2}\mathcal{I}_{0}(b_{i})+\epsilon_{i} Z_{\epsilon}(\xi_{i})\Gamma_{\epsilon}(b_{i})\}
\end{eqnarray}
where $W^{ad,s}$ are given in Equation (\ref{eq:3.18.3}) and we will use the ratio between $\epsilon_{e}$ and $\epsilon_{i}$ from Equation (\ref{eq:3.18.4}).
%%%%%%%%%%%%%%%%%%%%%%%%%%%%%%%%%%%%
\section{Results and discussion}
%%%%%%%%%%%%%%%%%%%%%%%%%%%%%%%%%%%%
We have derived a dispersion relation for drift waves driven by a density gradient in a shear-less slab geometry with constant magnetic field where the small deviation from a Maxwellian distribution is described by $\epsilon$. Here we will determine the quantitative effects on the real frequency and growth rate as a function of this deviation. We start by assuming that we have adiabatic electrons for which the dispersion Equation (\ref{eq:3.28}) is,
\begin{eqnarray}\label{eq:4.1}
2+\epsilon_{i} (2.35\;W^{ad,e}+ W^{ad,i})=\nonumber\\
-(\xi_{i}-\bar{\omega}_{*,i})\{(-\xi_{i}^{-1}-\frac{1}{2}\xi_{i}^{-3})e^{-b_{i}/2}\mathcal{I}_{0}(b_{i})+\epsilon_{i} Z_{\epsilon}(\xi_{i})\Gamma_{\epsilon}(b_{i})\}.
\end{eqnarray}
After rearranging the terms in the above equation we finally get the following relation for $\epsilon_{i}$:
\begin{eqnarray}\label{eq:4.1.1}
\epsilon_{i} =\frac{-2\xi_{i}^3+[\xi_{i}^3+0.5\xi_{i}-\bar{\omega}_{*,i}\xi_{i}^2-0.5\bar{\omega}_{*,i}]e^{-b_{i}/2}\mathcal{I}_{0}(b_{i})}{W^{ad,tot}\xi_{i}^3+(\bar{\omega}_{*,i}\xi_{i}^3-\xi_{i}^4)Z_{\epsilon}(\xi_{i})\Gamma_{\epsilon}(b_{i})}
\end{eqnarray}
where $W^{ad,tot}=2.35\;W^{ad,e}+ W^{ad,i}$. This relation gives the possible deviation of the equilibrium PDF from the Maxwellian PDF for a given plasma turbulence, i.e $\xi_{i}$. One has to remember that only positive values of $\mathbf{Re}[\epsilon]$ are physically meaningful.

\begin{figure}[tbp]
\begin{center}
\epsfig{figure=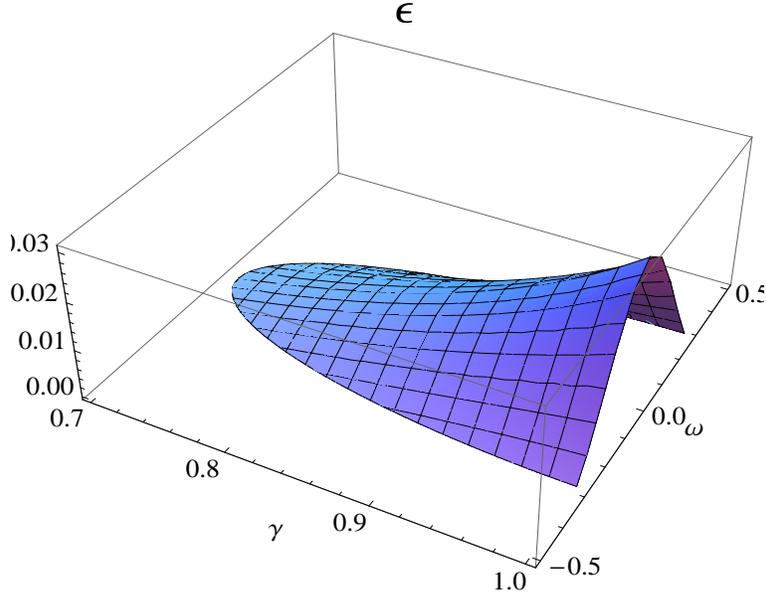, width=10cm,height=8cm,clip=}
\end{center}
\caption{$\epsilon$ as a function of $\omega$ and $\gamma$ where $\xi_{i}=\omega+i \gamma$. We have assumed $b_{i}=0.1$, $k_{\parallel}= 10^{-3}$ and $\bar{\omega}_{*,i}=-7.1\times 10^{2}$ with $d\;ln\;n/dx=1$.}
\label{fig1}
\end{figure}

Using the same plasma parameters as was used in Equations (\ref{eq:3.18.3}) and (\ref{eq:3.24},\ref{eq:3.26}) we compute the term $Z_{\epsilon}(\xi_{i})$, and from Equation (\ref{eq:3.22}) we get
\begin{eqnarray}\label{eq:4.2}
Z_{\epsilon}(\xi_{i})=\frac{V_{T,i}}{\sqrt{\pi}}Lim_{\sigma\rightarrow 0}\{\frac{1}{(-\xi_{i}-i\sigma)}\int_{-\infty}^{\infty}du\Phi(u)+\frac{1}{(-\xi_{i}-i\sigma)^3}\int_{-\infty}^{\infty}u^2du\Phi(u)\}\nonumber\\
=\frac{-6.5\times 10^{-9} - 3.8 \times 10^{-9} \xi_{i}^2}{\xi_{i}^3}.
\end{eqnarray}
Here, those integrations omitted resulted in zero contributions and rewriting Equation (\ref{eq:4.1}) by using these explicit values results in the expression for the deviation in Equation (\ref{eq:4.1.1}) we obtain
\begin{eqnarray}\label{eq:4.4}
\epsilon_{i} =\frac{-2\xi_{i}^3+[\xi_{i}^3+0.5\xi_{i}-\bar{\omega}_{*,i}\xi_{i}^2-0.5\bar{\omega}_{*,i}]e^{-b_{i}/2}\mathcal{I}_{0}(b_{i})}{66.3\xi_{i}^3-39.2\xi_{i}+39.2\bar{\omega}_{*,i}+23.0 \bar{\omega}_{*,i}\xi_{i}^2}
\end{eqnarray}
Figure \ref{fig1} shows $\epsilon_{i}$ from Equation (\ref{eq:4.4})  where $\xi_{i}=\omega+i \gamma$. Here, the values of $\omega, \gamma$ are normalized to $|k_{\parallel}|V_{T,i}$. We have assumed parameter values $b_{i}=0.1$, $k_{\parallel}=10^{-3}$ and $\bar{\omega}_{*,i}=-7.1\times 10^{2}$ with $d\;ln\;n/dx=1$. It is found that there is a threshold in the growth rate $\gamma$ close to $\gamma = 0.7$ and that increasing to $1.0$ only increases the deviation from a Maxwellian from 0 to 0.03. It should be noted that $\epsilon$ increases the excess kurtosis of the distribution function by a similar amount thus a quite small deviation from a Maxwellian can have a rather significant impact.

\begin{figure}[tbp]
\begin{center}
\epsfig{figure=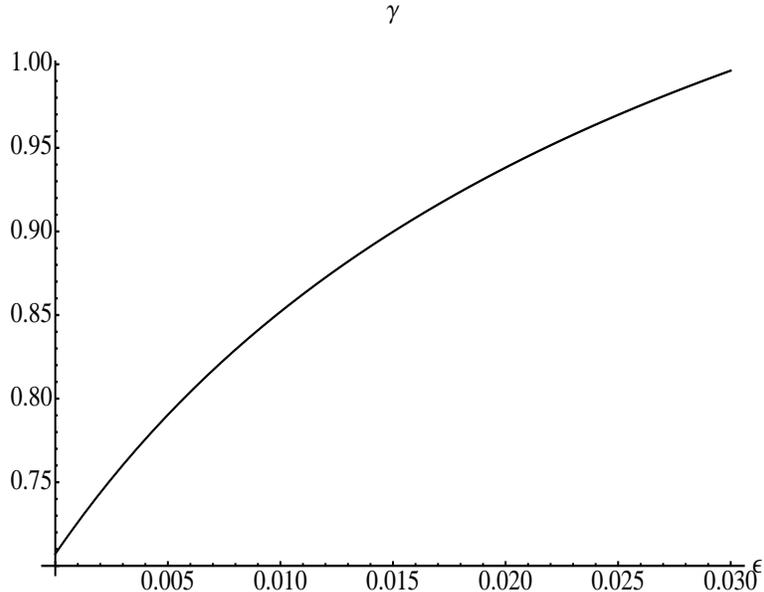, width=10cm,height=8cm,clip=}
\end{center}
\caption{$\gamma$ as a function of $\epsilon$. The same plasma parameters as in the figure \ref{fig1} are used.}
\label{fig2}
\end{figure}

In figure \ref{fig2}, the mode growth rate as a function of $\epsilon_{i}$ is shown. Note that in this figure the values of growth rate are the solutions of the Equation (\ref{eq:4.4}) for a given $\epsilon_{i}$ while in the figure \ref{fig1} we solve Equation (\ref{eq:4.4}) for $\epsilon$ at a given $\xi_{i}$. As the dispersion equation is of 3rd order in $\bar{\omega}$ three possible solutions exist, however we are only interested in the solutions with non-zero imaginary value, $\gamma>0$ corresponding to unstable situations. It is shown in figure \ref{fig2} that a deviation of $\epsilon_{i}=0.01$ yield an increase of about $20\%$ in the growth rate. Furthermore, the growth rate increases almost linearly with increasing $\epsilon_{i}$ and such an increase in the growth rate will lead to a significant increase in the level of anomalous flux.

%%%%%%%%%%%%%bild%%%%%%%%%%%%%%%%%%%%
%%%%%%%%%%%%%bild%%%%%%%%%%%%%%%%%%%%
%%%%%%%%%%%%%bild%%%%%%%%%%%%%%%%%%%%
%%%%%%%%%%%%%bild%%%%%%%%%%%%%%%%%%%%

In summary, we have derived a dispersion relation for density gradient driven linear drift waves including the effects coming from the inclusion of a fractional velocity derivative in the Fokker-Planck equation in the case of constant magnetic field and a shear-less slab geometry. The solutions of this Fokker-Planck equation are the alpha-stable distributions. It has not yet been shown that in a direct way one can derive the alpha-stable distribution function \cite{BalescuBook,montroll} from the classical form of collision operator \cite{Gatto}. One way may be to construct a new type of collisional operator by considering a fractal phase space and reformulate the collision operator on this new space. However, such a discussion is outside the scope of the present paper. Interestingly enough, we note that non-local effects are observed in non-linear collisionless fluid simulations of plasma turbulence where the non-local transport showing Levy features are induced by the interaction of the non-linear terms in the dynamical equations \cite{negrete2005}. The non-local features of non-linear fluid models are indicated by recent analytical theories using path-integral methods to derive probability density fucntions of fluxes \cite{anderson1}.

The fractional derivative is represented with the Fourier transform containing a fractional exponent that we are able to connect to the deviation from a Maxwellian distribution described by $\epsilon$. The characteristics of the plasma drift wave are fundamentally changed, i.e. the values of the growth-rate $\gamma$ and real frequency $\omega$ are significantly altered. A deviation from the Maxwellian distribution function alters the dispersion relation for the density gradient drift waves such that the growth rates are substantially increased and thereby may cause enhanced levels of transport.

{\em Acknowledgements}
The authors would like to thank professor T. F\"ul\"op for her helpful comments. This work was funded by the European Communities under Association Contract between EURATOM and
{\em Vetenskapsr{\aa}det}.

\end{document}